\documentclass[%
 aip,
cp,  
 amsmath,amssymb,
 reprint,%
]{revtex4-2}

\usepackage{graphicx}
\usepackage{dcolumn}
\usepackage{bm}

\usepackage[utf8]{inputenc}
\usepackage[T1]{fontenc}
\usepackage{mathptmx} 
\usepackage{xcolor,amsmath}%
\usepackage{ccaption} 

\begin{document}

\title{Effect of Geometric Confinement on the Stabilization of Skyrmions}

\author{H. Koibuchi} 
 \email{koibuchi@gm.ibaraki-ct.ac.jp; koibuchih@gmail.com}
\author{F. Kato}%
\affiliation{
  National Institute of Technology (KOSEN), Ibaraki College, Hitachinaka, Japan.
}
\author{S. El Hog}
\affiliation{%
Universit$\acute{\rm e}$ de Monastir (LMCN), Monastir, Tunisie
}
\author{G. Diguet}
 \email{gildas.diguet.d4@tohoku.ac.jp}
\affiliation{%
Micro System Integration Center, Tohoku University, Sendai, Japan
}%
\author{B. Ducharne}
\affiliation{%
INSA Lyon, Universite de Lyon, Villeurbanne Cedex, France
}%
\affiliation{ElyTMax, CNRS-Universite de Lyon-Tohoku University, Sendai, Japan}
\author{\\T. Uchimoto}
\affiliation{%
Institute of Fluid Science (IFS), Tohoku University, Sendai, Japan
}%
\affiliation{ElyTMax, CNRS-Universite de Lyon-Tohoku University, Sendai, Japan}
\author{H. T. Diep}
 \email{diep@cyu.fr}
\affiliation{%
CY Cergy Paris University, Cergy-Pontoise, France
}%

\date{\today} 

\begin{abstract}
In this paper, we study the stability of skyrmions (SKYs) caused by the geometric confinement (GC) effect observed in nano-domains in recent experiments, where SKYs appear only inside the boundary and is stable at the low magnetic field region. However, the mechanism of the GC effect is unclear for skyrmions.  We numerically find that this effect is not observed in the standard model, in which the Dzyaloshinskii-Moriya interaction (DMI) energy is discretized on lattice edges, while the effect is observed in a discrete model where the DMI energy is defined on lattice volume elements.  In the latter model, the DMI energy on the surfaces effectively becomes very small compared with that of the inside. From this observation, we study a model with zero DMI energy on the surfaces parallel to the external magnetic field and find that SKY is significantly stabilized in the sample in the low magnetic field region.
\end{abstract}

\maketitle

\section{\label{sec:intro}Introduction} 

Skyrmions (SKYs) are topologically stable spin configurations in chiral magnetic materials, and their stability is one of the main topics that should be studied in view of future technological applications \cite{Fert-etal-NatReview2017,Zhang-etal-JPhys2020,Gobel-etal-PhysRep2021}. An external magnetic field is a primary ingredient for stabilization \cite{Butenko-etal-PRB2010}, and mechanical stresses influencing shape deformation \cite{Shibata-etal-Natnanotech2015,Koretsune-etal-SCRep2015,ElHog-etal-PRB2021} are also known to enhance stability \cite{Butenko-etal-PRB2010,Seki-etal-PRB2017,Chacon-etal-PRL2015,Nii-etal-NatCom2015,ElHog-etal-RinP2022}. 

The geometric confinement (GC) effect \cite{Rohart-Thiaville-PRB2013} was observed in a nanostripe of the chiral magnet FeGe \cite{HDu-etal-NatCom2015}, and morphological changes in skyrmions under various  nanostripe widths were reported in  \cite{CJin-etal-NatCom2017}. 
Skyrmion bubbles in centrosymmetric magnets are also influenced by GC effects \cite{ZHou-etal-AcsNano2019}, where the applied magnetic field decreases with decreasing width of the nanostripe causing stabilization.

Recent experiments also reported that confined SKYs are stabilized  in a multilayered nanodomain  \cite{PHo-etal-PRAp2019}, and the confined SKYs in a nanodomain are further stabilized  under the presence of tensile strains \cite{YWang-etal-NatCom2020}. This stabilization mechanism is closely related to GC, and this GC effect is expected to come from the so-called surface effects, which generally increase with reducing size. 
However, the origin of this effect is not always clear in the case of skyrmions. 
 
To find the mechanism, we first compare the results of two different three-dimensional (3D) models, which are different in the definition of discrete energies corresponding to the ferromagnetic interaction (FMI) and Dzyaloshinskii-Moriya interaction (DMI). In one of the models, denoted by model 1,  both FMI and DMI energies are defined on edges or bonds, called 1-simplex, of tetrahedrons and almost uniformly distributed independent of the position of bonds. In the other model, denoted by model 2, FMI and  DMI energies are defined on tetrahedrons, called 3-simplex,  and for this reason,  the distributions of FMI and  DMI energies  on 1-simplex are not uniform but low (high) density on (in the inside) the boundary surface,  resulting from a surface effect. In summary, 
\begin{enumerate}
\item [] DMI energy is defined on the bonds    :  model 1,        
\item [] DMI energy is defined on the volume element  : model 2.
\end{enumerate}
The boundary condition is the free boundary in both models. 
Here, the surface effect in model 2 is characterized by the fact that bonds on the surface are surrounded or shared by the bulk volume only on one side or partly. Such a surface effect is not shared with model 1.  From the numerical data, we find that SKY states are confined (not confined)  inside the boundary in  model 2 (model 1). 

From this observation, we numerically study a simplified 1-simplex or edge model denoted by model 3 with zero DMI energy on the boundary surfaces parallel to the magnetic field.   The results of Monte Carlo simulations of model 3 indicate that the zero-DMI energy on the surfaces parallel to the magnetic field stabilizes skyrmions in a small magnetic field region, consistent with the reported experimental data in Ref. \cite{YWang-etal-NatCom2020}. This numerically observed stabilization is caused by a heterogeneous distribution of DMI  energy between the surface and the inside or bulk and explains the mechanism of the geometric confinement effect.  Anisotropic coupling constants were also confirmed to stabilize skyrmions \cite{Osorio-etal-PRB2019,SGao-etal-Nat2020,DAmoroso-etal-Nat2020,WCLi-etal-PhysScr2022}.

\section{Model}
\begin{figure}[h]
\centering{}\includegraphics[width=16.0cm]{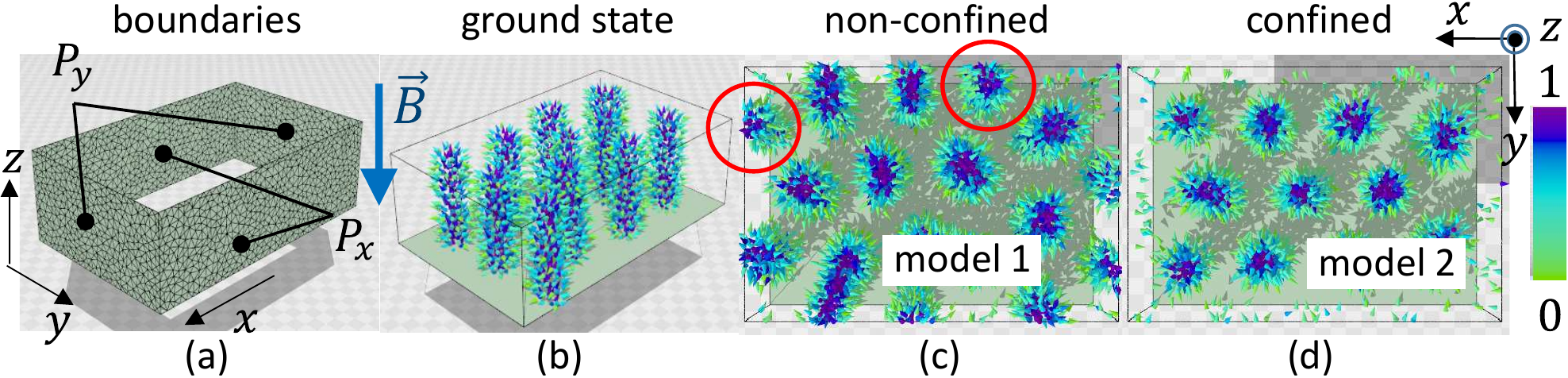}
\caption{(a) The boundary surfaces $P_x$ and $P_y$ parallel to the $z$ axis of a 3D tetrahedral lattice, (b) a ground state of Neel type SKYs, where the magnetic field is $\vec{B}\!=\!(0,0,-B)$, and snapshots of (c) non-confined SKYs of model 1 and (d) confined SKYs of model 2. Red circles in (c) indicate non-confined SKYs appearing on the boundary in model 1.  \label{fig-1}  }
\end{figure}

The boundary surfaces of a 3D lattice discretized by tetrahedrons are shown in Fig. \ref{fig-1}(a). This 3D lattice  is used to define two different models 1 and 2. Both models 1 and 2 are defined by the Hamiltonian composed of three different terms
\begin{eqnarray}
S=\lambda S_{\rm FM}+DS_{\rm DM}-S_{B}, \quad S_{B}=\sum_i \vec{\sigma}_i\cdot \vec{B},\quad  \vec{B}=(0,0,-B),
\label{total-Hamiltonian}
\end{eqnarray}
where Zeeman energy $S_{B}$ are common to both models. The FMI energy $S_{\rm FM}$  and DMI energy $S_{\rm DM}$ of models 1 and 2 are defined by
\begin{eqnarray}
\begin{split}
& S_{\rm FM}=\sum_{ij}\left(1-\vec{\sigma}_i\cdot\vec{\sigma}_j\right),  \quad S_{\rm DM}=\sum_{ij}\left(\vec{e}_{ij}\times\vec{e}_z\right)\cdot\left(\vec{\sigma}_i\times \vec{\sigma}_j\right), \quad({\rm model\; 1}),\\
&S_{\rm FM}=\sum_{ij}n_{ij}\left(1-\vec{\sigma}_i\cdot\vec{\sigma}_j\right),  \quad S_{\rm DM}=\sum_{ij}n_{ij}\left(\vec{e}_{ij}\times\vec{e}_z\right)\cdot\left(\vec{\sigma}_i\times \vec{\sigma}_j\right), \quad({\rm model\; 2}),\\
& n_{ij}=\bar{N}^{-1}\sum_{{\Delta}(ij)}1, \quad \bar{N}=\sum_{ij}\sum_{{\Delta}(ij)}1/\sum_{ij}1=6N_{\rm tet}/N_B\simeq 4.91.
\end{split}
\label{model-AB}
\end{eqnarray}
A free boundary condition is assumed on all boundaries for the variable $\vec{\sigma}_i(\in\! S^2\!:\!{\rm unit \;sphere})$, which denotes the spin variable at vertex $i$,  $\vec{e}_{ij}$ is the unit tangential vector from $i$ to $j$, and $\vec{e}_z\!=\!(0,0,1)$. In model 1, both $S_{\rm FM}$ and $S_{\rm DM}$ are defined by the sum $\sum_{ij}$ over bonds $ij$, whereas in model 2, these are defined by the sum $\sum_{ \Delta}\sum_{ij({\Delta})}$  over tetrahedrons ${\Delta}$ and the bonds ${ij({\Delta})}$ of ${\Delta}$. This summation $\sum_{\Delta}\sum_{ij({\Delta})}$ is a possible discretization of the volume integration of a continuous Hamiltonian such as $S_{\rm FM}=\frac{1}{2}\int  \sqrt{g}d^3x g^{ab}\frac{\partial \vec{\sigma}}{\partial x^a}\cdot \frac{\partial \vec{\sigma}}{\partial x^b}$. The $S_{\rm DM}$ in both models is the so-called  interfacial DMI corresponding to Neel type skyrmion (SKY) \cite{ERuff-etal-SciAdv2015,IKezsmarki-etal-NatMat2015,YFujimka-etal-PRB2017,YWu-etal-NatCom2020}. The constant $\bar{N}$ in the $S_{\rm FM}$ and $S_{\rm DM}$ of model 2 is defined by $\bar{N}\!=\!\sum_{ij}\sum_{{\Delta}(ij)}1/N_B$, where $\sum_{{\Delta}(ij)}1$ is the total number of tetrahedrons sharing the bond $ij$, and $N_B\!=\!\sum_{ij}1$ is the total number of bonds. 
This factor  $\bar{N}$ is the mean value of $\sum_{\Delta(ij)}1$. Using the relation $\sum_{ \Delta}\sum_{ij({\Delta})}1\!=\!\sum_{ij}\sum_{\Delta(ij)}1$, we obtain 
$\bar{N}\!=\!\sum_{ij}\sum_{{\Delta}(ij)}1/\sum_{ij}1\!=\!\sum_{{\Delta}}\sum_{ij(\Delta)}1/\sum_{ij}1\!=\!6N_{\rm tet}/N_B\!\simeq\!4.91$, where $N_{\rm tet}$ is the total number of tetrahedrons. In this expression, we use the fact that  tetrahedron has 6 bonds, and hence $\sum_{{\Delta}}\sum_{ij(\Delta)}1\!=\!\sum_{{\Delta}}6\!=\!6N_{\rm tet}$. 
 Because of this factor $\bar{N}$, the mean value of $n_{ij}$ is $\langle n_{ij}\rangle\!=\!\sum_{ij}n_{ij}/\sum_{ij}1\!=\!1$.  This relation $\langle n_{ij}\rangle\!=\!1$ is the reason why $\bar{N}$ is introduced. 
 Note that $n_{ij}$ depends on whether the bond $ij$ is on the surfaces or inside. This position dependence of $n_{ij}$ implies that the distribution of $S_{\rm DM}$ in model 2 is heterogeneous because the element $\left(\vec{e}_{ij}\!-\!\vec{e}_z\right)\cdot\left(\vec{\sigma}_i\times \vec{\sigma}_j\right)$ of $S_{\rm DM}$ is defined on the bond $ij$. 
 More detailed information on the Hamiltonian discretization will be reported elsewhere.  

A ground state of the SKY phase of model 2 is shown in Fig. \ref{fig-1}(b), where the magnetic field $ \vec{B}\!=\!(0,0,-B)$ is applied along the negative $z$ direction. Top views of snapshots obtained after $1\!\times\! 10^8$ Monte Carlo sweeps for models 1 and 2 are shown in Figs. \ref{fig-1}(c) and \ref{fig-1}(d), respectively. We find  that SKY states of model 1 appear even on the surfaces $P_x$ and $P_y$  parallel to the $z$ axis in Fig. \ref{fig-1}(c) while no SKY on the surfaces in Fig. \ref{fig-1}(d) of model 2. All SKY states of model 2 are confined inside the boundary. 

\begin{figure}[h]
\centering{}\includegraphics[width=11.5cm]{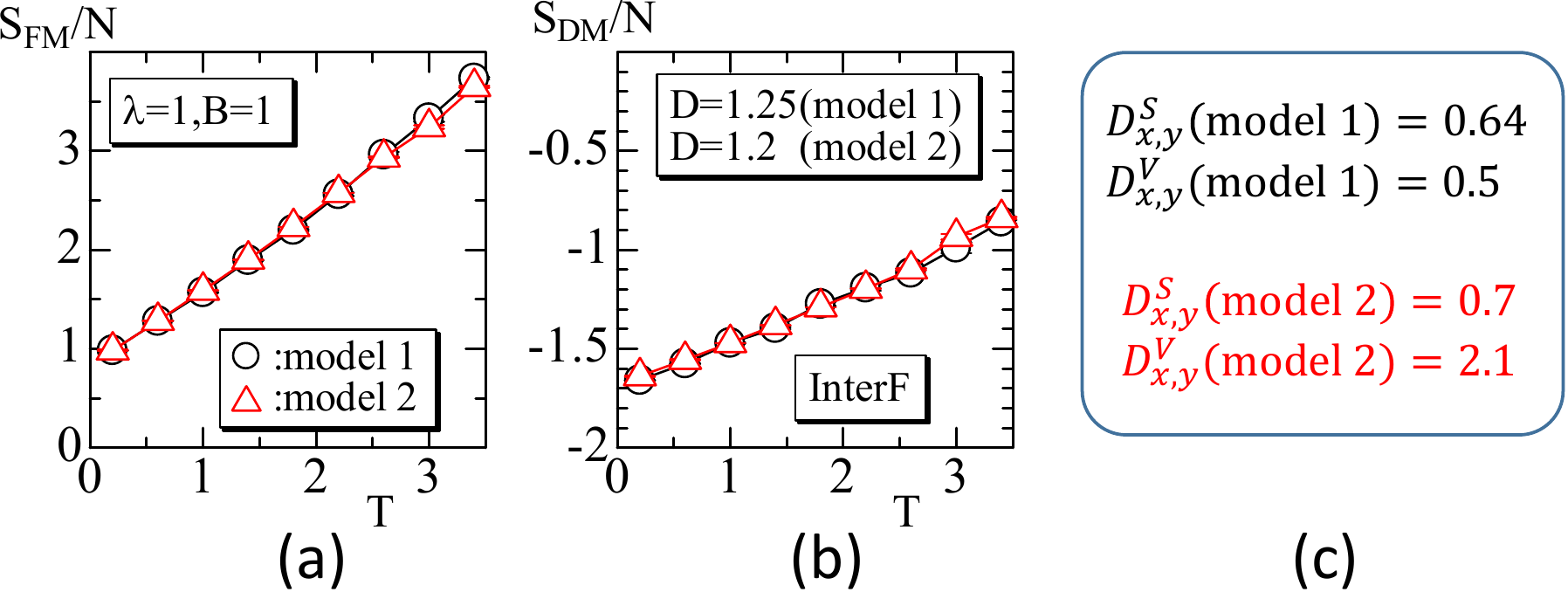}
\caption {(a) $S_{\rm FM}$ vs. $T$, (b)  $S_{\rm DM}$ vs. $T$ obtained with  $(\lambda,D,B)\!=\!(1,1.25,-1)$ for model 1 and  $(\lambda,D,B)\!=\!(1,1.2,-1)$ for model 2. The difference in the parameter sets comes from the difference in the discrete forms of models 1 and 2. (c) Comparison of effective coupling constants of surfaces and bulk defined by Eq. (\ref{DMI-Coeff}). $D_\mu^{\rm S}$ is comparable with $D_\mu^{\rm V}$ in model 1. In contrast,  $D_\mu^{\rm S}$ is three times smaller than  $D_\mu^{\rm V}$ in model 2. This large difference in the DMI coupling constants comes from a surface effect. ``Interf'' in (b) denotes the interfacial DMI model.
\label{fig-2} }
\end{figure}

The parameters $(\lambda,D,B)$ are fixed to $(\lambda,D,B)\!=\!(1,1.25,-1)$ for model 1 and  $(\lambda,D,B)\!=\!(1,1.2,-1)$ for model 2 so that $S_{\rm FM}/N$ and $S_{\rm DM}/N$ vs. $T$ are almost the same in both models as shown in Figs. \ref{fig-2}(a),(b), where $N$ is the total number of vertices.  For the comparison of the results of the two models, the obtained skyrmion configurations should be almost the same, and for this reason, we plot  $S_{\rm FM}/N$ and $S_{\rm DM}/N$ to check whether these are almost the same. 

To see a difference between the results of models 1 and 2, we define position-and-direction dependent constants as follows: 
\begin{eqnarray}
\begin{split}
&D_x^S=\frac{1}{\sum_{ij\in P_x}1}\sum_{ij\in P_y} n_{ij}|{e}_{ij}^{y}|,\quad
D_y^S=\frac{1}{\sum_{ij\in P_x}1}\sum_{ij\in P_y} n_{ij}|{e}_{ij}^{x}|\quad
D_\mu^V=\frac{1}{\sum_{ij\in V\setminus S}1}\sum_{ij\in V\setminus S} n_{ij}|{e}_{ij}^{\mu}|,\;(\mu=x,y),\\
&S\!=\!P_x\cup P_y\cup P_z,\quad  n_{ij}=\left\{ \begin{array}{@{\,}ll}
                 1&  ({\rm model\; 1}) \\
                 (1/\bar N)\sum_{{\Delta}(ij)}1, &  ({\rm model\; 2})
                  \end{array} 
                   \right., \quad  
 \bar N=\sum_{ij}\sum_{{\Delta}(ij)}1/ \sum_{ij}1.
\end{split}
\label{DMI-Coeff}
\end{eqnarray}
Note that $D_z^{S,V}\!=\!0$ because of the expression $({\vec{e}}_{ij}\times \vec{e}_z)\cdot(\vec{\sigma}_{i}\times\vec{\sigma}_{j})=({\vec{e}}_{ij}\times \vec{e}_z)^x(\vec{\sigma}_{i}\times\vec{\sigma}_{j})^x+({\vec{e}}_{ij}\times \vec{e}_z)^y(\vec{\sigma}_{i}\times\vec{\sigma}_{j})^y$ in $S_{DM}$. 
The sums $\sum_{ij}$ in $D_\mu ^S$ and $\sum_{\Delta}\sum_{ij({\Delta})}$ in $D_\mu ^V$  are calculated on surfaces and inside or bulk $V\backslash S\!=\!V\!-\!S$, where $S$ is defined by $S\!=\!P_x\cup P_y\cup P_z$, $V$ denotes the whole volume, and $P_z$ denotes the upside and downside surfaces.  The symbol $\bar {N}$ denotes the mean value of $\sum_{{\Delta}(ij)}1$ the total number of tetrahedrons sharing a bond $ij$.  Note that ${e}^{x(y)}_{ij}\!=\!0$ on $P_{y(x)}$. These quantities $D_\mu^{S,V}$  in Eq. (\ref{DMI-Coeff}) are considered to be effective DMI coupling constants on the surfaces $S$ and  in the bulk $V\backslash S$, because $S_{\rm DM}$ depends on $n_{ij}$ and $\vec{e}_{ij}$ and this dependence can be extracted to define the constants. The surfaces denoted by $P_z$ are not regarded as a boundary surface because SKY penetrates through them as in the bulk part.

The values of effective coupling constants are shown on Fig. \ref{fig-2}(c). We find that $D_\mu^{\rm S}$ is comparable with $D_\mu^{\rm V}$ in model 1. However,  $D_\mu^{\rm S}$ in model 2 is three-times smaller than $D_\mu^{\rm V}$ implying that the DMI energy on the surfaces $P_x$ and $P_y$ becomes very small compared with that inside the surfaces. A small DMI energy means that SKY is unstable on the surfaces in model 2, which is why SKY does not appear on or is confined inside the boundary. 

Here,  we should note that $D_\mu^S$ and $D_\mu^V$ for $n_{ij}\!=\!1$ (model 1) are directly calculable by two or three-dimensional integration, respectively, because the distribution of $\vec{e}_{ij}$ is considered to be random. Indeed, using the polar coordinates on the unit disk for $D_x^S$ for example, $D_x^S\!=\!(1/{\sum_{ij\in P_x}1})\sum_{ij\in P_x} |{e}_{ij}^{x}|$ corresponds to $D_x^S\!=\!(1/\int_0^{2\pi} \sin\theta d\theta)\int_0^{2\pi} \sin\theta |\cos\theta| d\theta\!=\!2/\pi\!\simeq\!0.64$, and we also have $D_y^S\!=\!(1/\int_0^{2\pi} \sin\theta d\theta)\int_0^{2\pi} \sin\theta |\sin\theta| d\theta\!=\!2/\pi\!\simeq\!0.64$. The quantities $D_\mu^V$ are also calculable using the polar coordinates inside the unit sphere. Indeed, 
$D_x^V\!=\!(1/{\sum_{ij\in V\setminus S}1})\sum_{ij\in V\setminus S} n_{ij}|{e}_{ij}^{x}|$ can be replaced by  $D_x^V\!=\!(1/\int_0^{2\pi} \int_0^{\pi} \sin \theta d\theta  d\phi)\int_0^{2\pi}  \int_0^{\pi} \sin \theta |\cos\theta\cos\phi| d\theta d\phi\!=\!1/2$. The other quantity $D_y^V$ is also calculated to be $D_y^V\!=\!1/2$. These values obtained by integrations are almost the same as those shown in Fig. \ref{fig-2}(c) implying that the distribution of $\vec{e}_{ij}$ is almost uniformly random.

The results obtained by models 1 and 2 indicate that the small DMI energy on the surface $P_{xy}\!=\!P_x\cup P_y$ confines and stabilizes the bulk SKY inside the boundary. From this fact, we study model 3 defined by 
\begin{eqnarray}
S_{\rm FM}=\sum_{ij}\left(1-\vec{\sigma}_i\cdot\vec{\sigma}_j\right), \quad S_{\rm DM}=\sum_{ij}\Gamma_{ij}\left(\vec{e}_{ij}\times\vec{e}_z\right)\cdot\left(\vec{\sigma}_i\times \vec{\sigma}_j\right), \quad
\Gamma_{ij}=\left\{ \begin{array}{@{\,}ll}
                 0&  (ij \in P_{xy})\\
                 1 &  (ij \in V\backslash P_{xy}) 
                  \end{array} 
                   \right., \quad({\rm model\; 3}),    
\label{model-C}
\end{eqnarray}
where $\Gamma_{ij}$ makes $S_{\rm DM}$ zero on the boundary surface $P_{xy}\!=\!P_x\cup P_y$ parallel to the magnetic field $\vec{B}\!=\!(0,0,-B)$. Thus, model 3 is identical to model 1 except that DMI energy is fixed to zero on the boundary surface $P_{xy}$. The difference between models 1 and 3 is only due to the inclusion of $\Gamma_{ij}$ in $S_{\rm DM}$ of model 3.

\section{Results}
\begin{figure}[!h]
\begin{minipage}{0.27\linewidth}
\caption{(a) $BT$ phase diagram and (b)--(h) snapshots of model 1. C-sky and nc-sky denote confined and non-confined SKYs, and Skst denotes an intermediate state between SKY and stripe, and skfe is an intermediate state between SKY and ferromagnetic states. For the region $|B|\geq 1.2$, SKY states are confined inside the boundary like (b), while SKY states appears on the boundary and are non-confined like those in (c)--(e) in the region $|B|\leq 1$.    \label{fig-3}}
\end{minipage}
\hspace{0.01\linewidth}
\begin{minipage}{0.70\linewidth}
\centering{}\includegraphics[width=10.5cm]{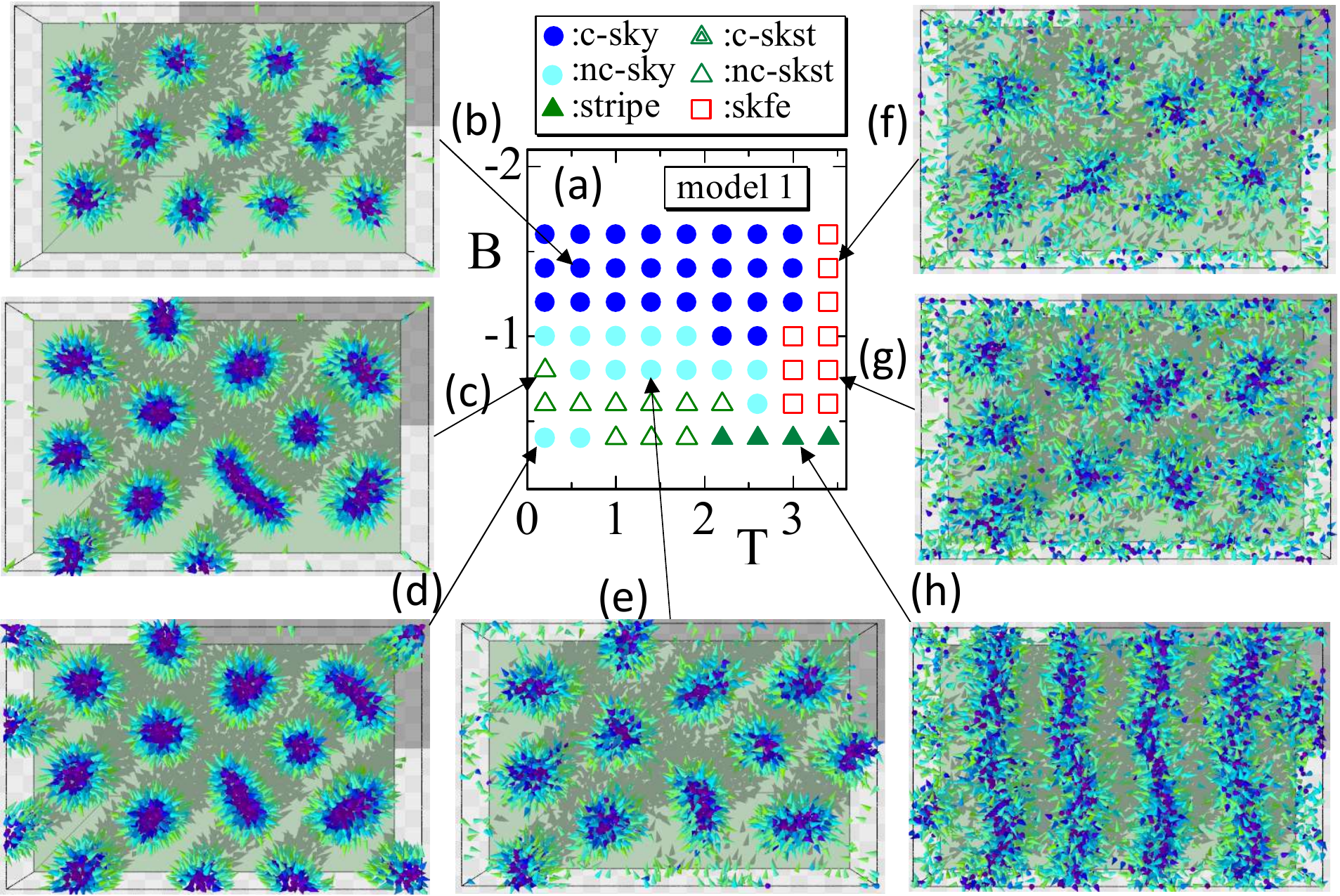}
\end{minipage}
\end{figure}

\begin{figure}[!h]
\begin{minipage}{0.27\linewidth}
\caption{(a) $BT$ phase diagram and (b)--(h) snapshots of model 3. SKY states are confined inside the boundary as shown in (b)--(e) not only in large $|B|$ but also in small  $|B|$ region.  SKY-stripe state shown in (d) is also a confined state because neither SKY state nor stripe state touches the boundary, and hence this is denoted by confined SKY-stripe state (c-skst). Two of four stripe states in (h) are also confined. \label{fig-4}}
\end{minipage}
\hspace{0.01\linewidth}
\begin{minipage}{0.70\linewidth}
\centering{}\includegraphics[width=10.5cm]{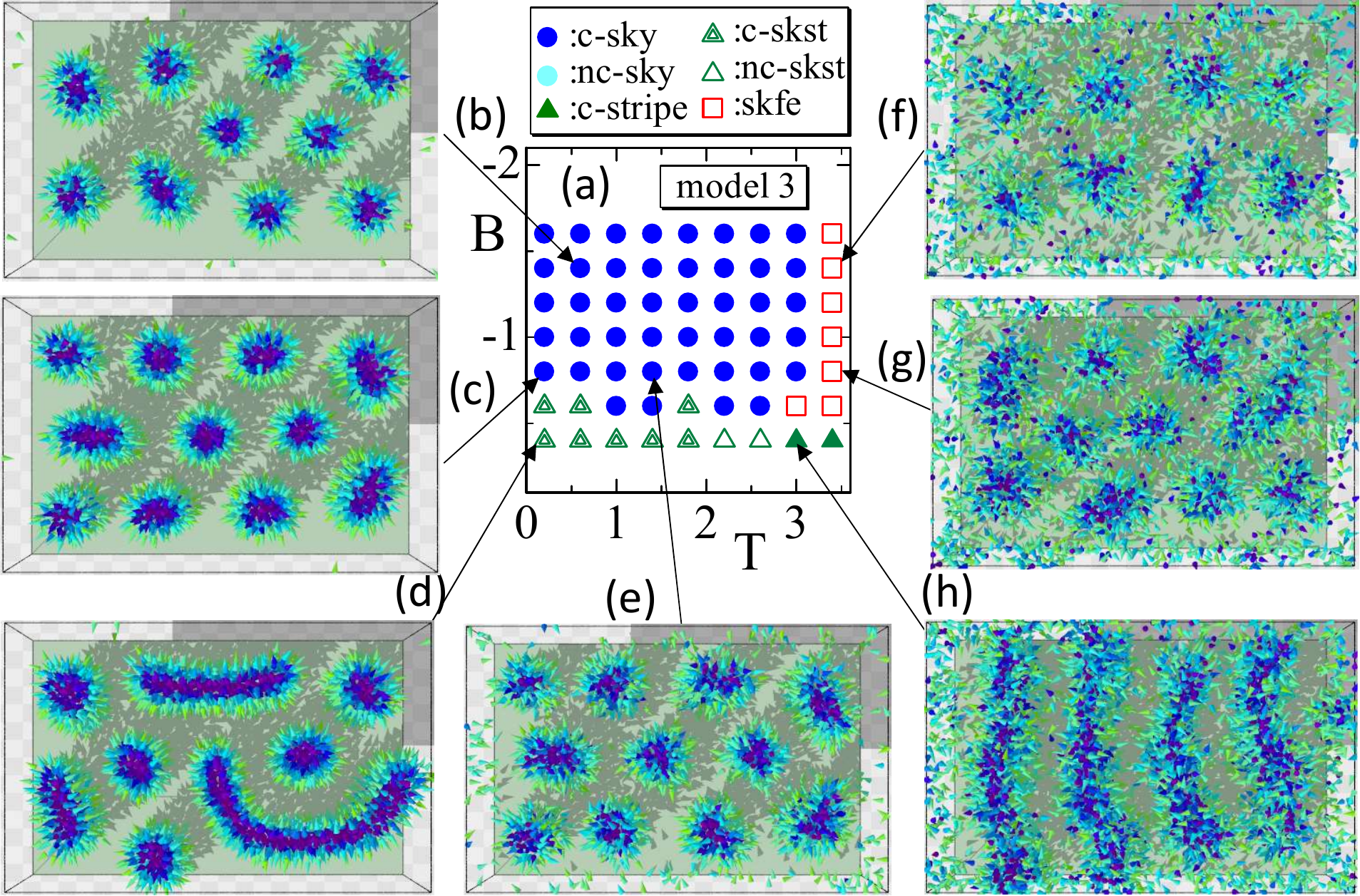}
\end{minipage}
\end{figure}

The $BT$ phase diagrams of models 1 and 3 are shown in Figs. \ref{fig-3} and \ref{fig-4}, where the constants $\lambda$ and $D$ are fixed to $(\lambda,D)\!=\!(1,1.25)$  in both models 1 and 3. 
Symbols (\textcolor{blue}{\Large{$\bullet$}}) and (\textcolor{-red}{\Large{$\bullet$}}) denote confined skyrmion (c-sky) and nonconfined skyrmion (nc-sky), respectively. 
We find that the SKY configurations of model 1 in Fig. \ref{fig-3} are nonconfined at low $|B|$ region compared with those of model 3 in Fig. \ref{fig-4}. The confined skyrmion in Fig. \ref{fig-3}(a) for model 1 appears only for the region $|B|\!\geq\! 1$. In contrast, in the case of model 3 in Fig.\ref{fig-4}(a)  c-sky appears for $|B|\!\geq\! 0.6$, and hence the area of c-sky in the $BT$ diagram of model 3 is expanded to lower values of $|B|$  compared with that of model 1. Moreover, an intermediate phase between stripe and skyrmion (skst) also turns from nonconfined skst (nc-skst) to confined skst (c-skst) or c-sky in the low $|B|$ region  if the model changes from  1 to 3.  We should note that skyrmions in model 1 are also confined in the large $|B|$ region. Therefore, the surface effect implemented by $\Gamma_{ij}$ or zero-DMI  in model 3 effectively increases $|B|$, which is compatible with the reported experimental data in Ref. \cite{YWang-etal-NatCom2020}.

Note  that model 3 is also an edge model similar to model 1, in which the discrete Hamiltonian is defined on edges or bonds of the lattice. Note also that model 2 is a volume model as described above.


\begin{figure}[h!]
\centering{}\includegraphics[width=14.5cm]{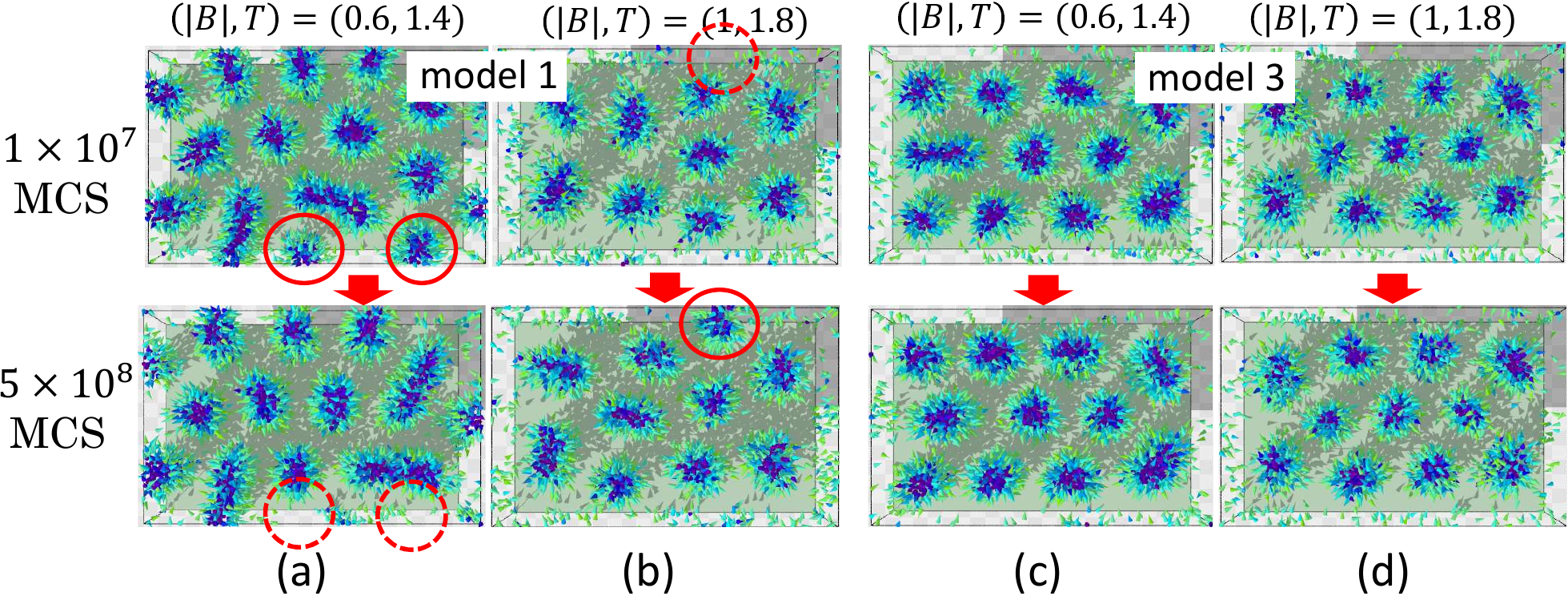}
\caption {Snapshots of model 1 obtained at (a) $(|B|,T)\!=\!(0.6,1.4)$, (b) $(|B|,T)\!=\!(1,1.8)$, and snapshots of model 3 obtained at (c) $(|B|,T)\!=\!(0.6,1.4)$, (d) $(|B|,T)\!=\!(1,1.8)$. $\Lambda$ and $D$ are fixed to $(\lambda,D)\!=\!(1,1.25)$  in both models 1 and 3.  Snapshots in the upper (lower) row are obtained at  $1\times 10^7$ ($5\times 10^8$) MCS.  Red circles show that SKY configurations (a) disappear and (b) appear in the lower row, indicating that the total number of SKYs decreases and increases, respectively and that nonconfined skyrmions are unstable.
\label{fig-5} }
\end{figure}

Next, we show that the confined SKY phase is stable in both models 1 and 3, and the nonconfined SKY phase observed in model 1 is unstable. Snapshots of the SKY phase of models 1 and 3 are shown in Figs. \ref{fig-5}(a),(b) and \ref{fig-5}(c),(d). Snapshots in the upper row are obtained at $1\times 10^7$ Monte Carlo sweep (MCS), while in the lower row, they are obtained at $5\times 10^8$ MCS, which is sufficiently larger than $1\times 10^7$ of the upper row. We find from Fig.  \ref{fig-5}(a) that nonconfined SKY (nc-sky) is unstable at least for low $|B|$ region. Indeed, some of the SKY configurations touching the boundary shrink or disappear as indicated by the red circles. A confined SKY (c-sky) phase in Fig. \ref{fig-5}(b) at relatively large $|B|(=\!1)$ is also unstable due to an emergence of SKY on the boundary $P_x$. Because of this emergence, the total number of SKYs is increased by one. From this observation, we find that stability/instability is closely connected to the creation/annihilation mechanism.  Snapshots of model 3 shown in Figs. \ref{fig-5}(c),(d) are both in the c-sky phase and stable even at relatively small $|B|(=\!0.6)$. Thus, we find that the c-sky phase is stable and the nc-sky phase is not always stable, and this observation is independent of whether the model is 1 or 3.

\section{Summary and conclusion}
The effect of a geometric confinement on the skyrmions (SKYs) is numerically studied on a 3D plate  tetrahedral-lattice using two different discrete models 1 and 2; the DMI energy of model 1 is uniformly distributed on the bonds independent of their position,  and the DMI energy of model 2  is  nonuniformly distributed on and inside the surfaces.  For this difference in models 1 and 2,  the surface DMI coefficient  effectively becomes three times smaller than the bulk DMI coefficient.  From the numerical results of these two models, we find  that SKY configurations in model 1 are not always confined inside the boundaries parallel to the magnetic field, while SKYs in model 2 are confined inside the boundary. For this reason,  we fix DMI coefficient zero in model 1 on the surfaces parallel to the magnetic field,  and find that no SKY state appears on the boundary, and the confined SKY states are significantly stabilized. 

A combination of the GC effect and mechanical strains reported in \cite{YWang-etal-NatCom2020} for the stabilization of skyrmions is interesting from a theoretical point of view.  The confinement effect for Bloch-type SKY is also interesting, and further numerical studies are necessary.

\begin{acknowledgments}
This work is supported in part by JSPS Grant-in-Aid for Scientific Research 19KK0095. Numerical simulations were performed on the Supercomputer system "AFI-NITY" at the Advanced Fluid Information Research Center, Institute of Fluid Science, Tohoku University.
\end{acknowledgments}


\nocite{*}
\bibliography{aipsamp}

\begin{thebibliography}{9}
\bibitem{Fert-etal-NatReview2017}
 A. Fert, N. Reyren and V. Cros,  \textit{Nature Reviews} {\bf 2} 17031 (2017).

\bibitem{Zhang-etal-JPhys2020}
X. Zhang , Y. Zhou, K. M. Song, T.E. Park, J. Xia, M. Ezawa, X. Liu, W. Zhao, G. Zhao and S. Woo, J. Phys.: Condens. Matter {\bf 32} 143001  (2020).

\bibitem{Gobel-etal-PhysRep2021}
B$\ddot{\rm o}$rge G$\ddot{\rm o}$bel, I. Mertig, O. A. Tretiakov, Phys. Rep. {\bf 895} 1-28,  (2021).

%
\bibitem{Butenko-etal-PRB2010}
 A. B. Butenko, A. A. Leonov, U. K.  R$\ddot{\rm o}$ssler, and A. N. Bogdanov, Phys. Rev. B \textbf{82}, 052403 (2010).



\bibitem{Shibata-etal-Natnanotech2015}
 K. Shibata, J. Iwasaki, N. Kanazawa, S. Aizawa,  T. Tanigaki, M. Shirai, T. Nakajima, M. Kubota,
 M. Kawasaki, H. S. Park, D. Shindo, N. Nagaosa and Y. Tokura 2015 \textit{Nature
Nanotech.} {\bf 10} 589 

\bibitem{Koretsune-etal-SCRep2015}
 T. Koretsune, N. Nagaosa and R. Arita, \textit{Scientific Reports} {\bf 75} 13302 (2015).

\bibitem{ElHog-etal-PRB2021}
 S. El Hog, F. Kato, H. Koibuchi and H. T. Diep, 
\textit{Phys. Rev. B} {\bf 104}, 024402(1-17) (2021), DOI: 10.1103/PhysRevB.104.024402


\bibitem{Seki-etal-PRB2017}
S. Seki, Y. Okamura, K. Shibata, R. Takagi, N. D. Khanh, F. Kagawa, T. Arima, and Y. Tokura,  Phys.Rev. B \textbf{96}, 220404(R) (2017).

\bibitem{Chacon-etal-PRL2015}
 A. Charcon, A. Bauer, T. Adams, F. Rucker, G. Brandle, R. Georgii, M. Garst, and C. Pfleidere,  
\textit{Phys. Rev. Lett.} {\bf 115},  267202 (2015), DOI: 10.1103/PhysRevLett.115.267202

\bibitem{Nii-etal-NatCom2015} Nii Y, Nakajima T, Kikkawa A, 
Y. Yamasaki, K.  Ohishi,   J. Suzuki,  Y. Taguchi, T. Arima, Y. Tokura and  Y. Iwasa, \textit{Nature Comm.} {\bf 6}, 9539  (2015).


\bibitem{ElHog-etal-RinP2022}
 S. El Hog, F. Kato F, S. Hongo, H. Koibuchi, G. Diguet, T. Uchimoto  and H.T. Diep, 
\textit{Results in Phys.}  https://doi.org/10.1016/j.rinp.2022.105578.

\bibitem{Rohart-Thiaville-PRB2013} 
S. Rohart and A. Thiaville, 
\textit{Phys. Rev. B } {\bf 88}, 184422 (2013), 
https://doi.org/10.1103/PhysRevB.88.184422

\bibitem{HDu-etal-NatCom2015} 
H. Du, R. Che, L. Kong X. Zhao, C. Jin, C. Wang, J. Yang,
W. Ning, R. Li, C. Jin, X. Chen, J. Zang, Y. Zhang and M. Tian, 
\textit{Nature Comm. } {\bf 6}, 8504 (2015), 
DOI: 10.1038/ncomms9504

\bibitem{CJin-etal-NatCom2017} 
C. Jin, Zi-An Li, A. Kov${\rm\acute{a}}$cs, J. Caron, F. Zheng, F. N. Rybakov,
N. S. Kiselev, H. Du, S. Bl${\rm\ddot{u}}$gel, M. Tian, Y. Z., M. Farle
and Rafal E. Dunin-Borkowski, 
\textit{Nature Comm. } {\bf 8}, 15569 (2017), 
DOI: 10.1038/ncomms15569

\bibitem{ZHou-etal-AcsNano2019} 
Z. Hou, Q. Zhang, G. Xu, S. Zhang, C. Gong, B. Ding, H. Li,
F. Xu, Y. Yao, E. Liu, G. Wu, X. Zhang, and W. Wa,
\textit{ACS Nano. } {\bf 13}, 922-929 (2019), 
DOI:: 10.1021/acsnano.8b09689

\bibitem{PHo-etal-PRAp2019} 
P. Ho, A. K.C. Tan, S. G., A.L. G. Oyarce, M. Raju, L.S. Huang,
A. Soumyanarayanan, and C. Panagopoulos,
  \textit{Phys. Rev. Appl.} {\bf 11}, 024064 (2019), 
DOI: 10.1103/PhysRevApplied.11.024064.

\bibitem{YWang-etal-NatCom2020} 
Y.Wang, L. Wang, J. Xia, Z. Lai, G. T. X. Zhang, Z. Hou,
X. Gao, W. Mi, C. Feng, M. Zeng, G. Zhou, G. Yu, G. Wu,
Y. Zhou, W. Wang, X. Zhang and J. Liu
\textit{Nature Comm. } {\bf 11}, 3577 (2020), 
https://doi.org/10.1038/s41467-020-17354-7.

\bibitem{Osorio-etal-PRB2019}
S. A. Osorio, M. B. Sturla, H. D. Rosales, and D. C. Cabra
Phys. Rev. B {\bf 100}, 220404(R) (2019)
https://doi.org/10.1103/PhysRevB.100.220404

\bibitem{SGao-etal-Nat2020}
S. Gao, H. D. Rosales, F. A. G. Albarrac$\acute{\rm i}$n, V. Tsurkan, G. Kaur, T. Fennell, P. Steffens, M. Boehm, P. $\check{\rm C}$erm$\acute{\rm a}$k, A. Schneidewind, E. Ressouche, D. C. Cabra, C. R$\ddot{\rm u}$egg and O. Zaharko,
 Nature \textbf{586}, 37-41 (2020),
 https://doi.org/10.1038/s41586-020-2716-8

\bibitem{DAmoroso-etal-Nat2020}
D. Amoroso, P. Barone  and S. Picozzi,
 Nature Comm. \textbf{11}, 5784 (2020)
 https://doi.org/10.1038/s41467-020-19535-w

\bibitem{WCLi-etal-PhysScr2022}
W.C. Li, Z.Q. Liu, J.Y. Chen, D. Xie, X.W. Yao and Z.X. Deng, Phys. Scr. {\bf 97} 085818 (2022),
 https://doi.org/10.1088/1402-4896/ac8121


\bibitem{ERuff-etal-SciAdv2015}
E. Ruff, S. Widmann, P. Lunkenheimer, V. Tsurkan, S. Bord$\acute{\rm a}$ccs,
I. K$\acute{\rm e}$zsm$\acute{\rm a}$rki, A. Loidl,
Sci. Adv. {\bf 1}, e1500916 (2015),
DOI: 10.1126/sciadv.1500916

\bibitem{IKezsmarki-etal-NatMat2015}
I. K$\acute{\rm e}$zsm$\acute{\rm a}$rki, S. Bord$\acute{\rm a}$cs, P. Milde, E. Neuber, L. M. Eng, J. S. White, H. M. Ronnow, C. D. Dewhurst, M. Mochizuki, K. Yanai, H. Nakamura, D. Ehlers, V. Tsurkan and A. Loidl,
Nature Matter {\bf 14}, 1116-1122 (2015),
https://doi.org/10.1038/nmat4402

\bibitem{YFujimka-etal-PRB2017}
Y. Fujima, N. Abe, Y. Tokunaga, and T. Arima, 
Phys. Rev. B {\bf 95}, 180410(R) (2017)
https://doi.org/10.1103/PhysRevB.95.180410

\bibitem{YWu-etal-NatCom2020}
Y. Wu, S. Zhang, J. Zhang, W. Wang, Y.L. Zhu, J. Hu, G. Yin,
K. Wong, C. Fang, C. Wan, X. Han, Q. Shao, T. Taniguchi, K. Watanabe,
J. Zang, Z. Mao, X. Zhang and K.L. Wang, 
Nature Comm. {\bf 11}, 3860 (2020)
https://doi.org/10.1038/s41467-020-17566-x

\end{thebibliography}


\end{document}